\documentclass[prl,twocolumn,showpacs,amsmath,amssymb,noshowpacs]{revtex4}

\usepackage{graphicx}% Include figure files
\usepackage{dcolumn}% Align table columns on decimal point
\usepackage{bm}% bold math
\usepackage{color}

\begin{document}

%%% article title
\title{Electron-electron scattering and nonequilibrium noise in Sharvin contacts}

%%% author(s) ( + e-mail)
\author{K.\,E.\,Nagaev\thanks{e-mail: nag@cplire.ru},
        T.\,V.\,Krishtop,
        N.\,Yu.\,Sergeeva }

%%% author's address(es)
\address{%$^+$
  Institute of Radioengineering and Electronics, Moscow, 125009 Russia
  %\\~\\
  %$^*$Moscow Institute of Physics and Technology, Dolgoprudny,  141700 Russia
}

%%% dates of submition & resubmition (if submitted once, second argument is 

%%% abstract
\begin{abstract}
We consider wide ballistic microcontacts with electron-electron scattering in the leads and calculate electric noise and nonlinear conductance in them.
Due to a restricted geometry the collisions of electrons result in a shot noise even though they conserve the total momentum of electrons.
We obtain the noise and the conductivity for arbitrary relations between voltage $V$ and temperature $T$.
The positive inelastic correction to the Sharvin conductance is proportional to $T$ at low voltages $eV \ll T$, and to $|V|$ at high voltages.
At low voltages the noise is defined by the Nyquist relation and at high voltages the noise is related with the inelastic correction to the current by the Shottky formula $S_{in} = 2e\, I_{in}$.
\end{abstract}

\maketitle

\newcommand{\bn}{{\bf n}}
\newcommand{\bp}{{\bf p}}
\newcommand{\br}{{\bf r}}
\newcommand{\bR}{{\bf R}}
\newcommand{\bk}{{\bf k}}
\newcommand{\bv}{{\bf v}}
\newcommand{\wk}{\omega_{\bf k}}
\newcommand{\nk}{n_{\bf k}}
\newcommand{\brho}{\boldsymbol{\rho}}
\newcommand{\eps}{\varepsilon}
\newcommand{\la}{\langle}
\newcommand{\ra}{\rangle}
\newcommand{\be}{\begin{eqnarray}}
\newcommand{\ee}{\end{eqnarray}}
\newcommand{\mb}{\begin{multline}}
\newcommand{\me}{\end{multline}}
\newcommand{\intl}{\int\limits_{-\infty}^{\infty}}
\newcommand{\dE}{\delta{\cal E}^{ext}}
\newcommand{\SE}{S_{\cal E}^{ext}}
\newcommand{\dsp}{\displaystyle}
\newcommand{\phit}{\varphi_{\tau}}
\newcommand{\p}{\varphi}

Nonequilibrium electric noise is observed in most mesoscopic systems.
It depends on the conduction mechanism
and is more sensitive to the effects of electron-electron interactions than the average conductance ~\cite{Blanter99}.
In this article we are concerned with  Sharvin-type ballistic contacts.
In the absence of scattering near the contact, all the relaxation processes that lead to
dissipation and a finite resistance of the contact take place deep in the leads, where the electron
distribution is almost equilibrium. As the motion of electrons in the nonequilibrium region near the contact
is purely deterministic, the noise does not depend on the voltage and is specified by the Nyquist relation involving the equilibrium Sharvin conductance.
If any impurities are present in the contact, this results in a positive contribution to the resistance and in a shot noise, which is proportional to the current. Unlike the impurity scattering, electron-electron collisions do not contribute to the resistivity of a homogeneous conductor with a parabolic spectrum because they conserve the total momentum of electrons.
However very recently, it was shown both experimentally \cite{Renard08} and theoretically \cite{Nagaev08} that
electron--electron scattering may result in a negative correction to the resistance of wide ballistic
contacts. Therefore it is of interest to calculate the voltage-dependent electric noise in them and to find
out whether collisions of electrons result in a shot noise like impurity scattering.

Effects of electron-electron interaction on the shot noise have been extensively studied in the past for
contacts with imperfect transmission. More than a decade ago, they were considered semiclassically for diffusive multichannel microbridges~\cite{Nagaev95}.
More recently, a number of authors considered interaction effects in the shot noise of microstructures by
modeling them as conducting quantum dots that were either in the Kondo \cite{Mora08,Gogolin06} or Coulomb-blockade \cite{Loss00,Bagrets05} regime. The
interaction was assumed to take place between electrons in localized states on these dots. Naturally,
this interaction strongly differs from that in the bulk of the conductor. However our recent results show
that even collisions of electrons far from the contact affect the average current and hence may cause its
fluctuations.

To calculate the noise, we use the semiclassical Boltzmann\--Langevin method \cite{Kogan}. Previously, Kulik and
Omelyanchuk used a similar approach to calculate electric noise in Sharvin contacts caused by
electron-phonon scattering in the zero-temperature limit \cite{Kulik84}. Here we extend this approach to arbitrary temperatures.

We adopt a model of a ballistic contact similar to that of Kulik {\it et al.} \cite{Kulik77} for
the case of electron--phonon scattering. Consider two 2D electron gases separated by a thin impenetrable
barrier with a gap of width $2a$. We assume that $a$ is much larger than the Fermi wavelength and the
screening radius but much smaller than both elastic and inelastic mean free path of electrons. The distribution functions of electrons on both sides of the insulator obey the Boltzmann equation
\be
 \frac{\partial f}{\partial t}
 +
 \bv\,\frac{\partial f}{\partial\br}
 +
 e{\bf E}\,\frac{\partial f}{\partial\bp}
 =
 \hat{I}_{ee},
 \label{Boltz}
\ee
where ${\bf E} = -\nabla\varphi$ is the electric field. The electron--electron
collision integral in this equation is given by
\begin{align}
 I_{ee} = \frac{1}{2}\sum_{\bp'\bk\bk'}%\sum_{\bk}\sum_{\bk'}
 \Bigl[J(\bp'\bk'\to\bp\bk) - J(\bp\bk\to\bp'\bk')\Bigr],
 \label{I_ee,J}
\end{align}
where
\be
 J(\bp\bk\to\bp'\bk') = W(\bp\bk|\bp'\bk')\,f(\bp)\,f(\bk)  \nonumber\\
 \times
 [1-f(\bp')]\,[1-f(\bk')]
\label{J}
\ee
and $W(\bp\bk|\bp'\bk') = 8\pi^2\alpha_{ee}\nu^{-2}\, V_{vol}^{-3}\,\delta(\eps_{\bp} + \eps_{\bk} - \eps_{\bp'} - \eps_{\bk'})\,\delta(\bp + \bk - \bp' - \bk')$ is the probability of a transition from the state $(\bp, \bk)$ to the state $(\bp', \bk')$. Here $\alpha_{ee}$ is a dimensionless parameter of electron-electron scattering and
$\nu = m/\pi$ is the two-dimensional density of states.
Equation (\ref{Boltz}) should be solved together with the Poisson equation for the electric potential $\varphi$.
It is possible to avoid solving the latter  using the condition $E_F \gg{\rm max}(eV, T)$ \cite{Kulik84},
which means that in the absence of collisions, electrons near the Fermi surface just move along straight lines. This condition allows us to set ${\bf v} = v_F{\bf p}/p$ and remove the term with electric field from Eq. (\ref{Boltz}).

Now we should specify the boundary conditions to calculate the distribution functions using Eq. (\ref{Boltz}).
We set $f(\bp) = f_0(\eps_{\bp})$ and $\p=\pm V/2$ far from the gap in the left and right half-planes.

If collisions are neglected, $f(\bp, \br)$ depends solely on whether the electron trajectory originates from gap or not. It is convenient to use a notion of the angular domain $\Omega_{in}(\br)$ that contains all the momenta of electrons that came to point
$\br$ from the contact. In terms of this domain, the zero approximation distribution function is
{\setlength{\arraycolsep}{2pt}
\be
 f^{(0)}_{L,R}(\bp, \br) =
 \left\{
  \begin{array}{ll}
   f_0(\eps_{\bp} + e\p(\br) \mp eV/2), & \bp \notin \Omega_{in}(\br) \\
   f_0(\eps_{\bp} + e\p(\br) \pm eV/2), & \bp \in    \Omega_{in}(\br)
  \end{array}
 \right.
 \label{f_zero}
\ee}
for the electrons in left (upper sign) and right (lower sign) half-spaces, respectively.

The current through the contact is given by
\be
I = e \int\limits_{-a}^{a}{ d\rho \int {\frac{d^2p}{(2\pi)^2} v_{\perp} f(\bp, {\boldsymbol \rho}) }}.
\label{current}
\ee
where $v_{\perp}$ is the component of $\bf {v}$ normal to the insulator and
vector ${\boldsymbol \rho} = \bf{e}_{\parallel} \rho$ labels points within the gap in the plane of insulator.
Substituting expressions (\ref{f_zero}) into Eq. (\ref{current}) results in the well known expression for the Sharvin conductance
$G_0 = e^2 p_F a/\pi^2$.
The first-order correction in scattering to $G_0$ can be calculated by substituting $f(\bp, \br)$ from Eq. (\ref{Boltz}) into Eq. (\ref{current})
\be
 I_{in} =
 e\int\limits_{-a}^{a} d\rho
 \int \frac{d^2p}{(2\pi)^2}\, v_{\perp}
 \int\limits_0^{\infty} d\tau\, I_{ee}\{f^{(0)}(\bp(\tau),\br(\tau))\}.
 \label{dI}
\ee
Here $\tau$ is the time of travel to point $\boldsymbol \rho$ along the trajectory.

The collision integral (\ref{I_ee,J}) involves four electron momenta $\bp$, $\bk$, $\bp'$, and $\bk'$.
If none of electrons with these momenta crosses the gap (i.e. falls within $\Omega_{in}(\br)$), a substitution of distribution functions (\ref{f_zero}) into (\ref{I_ee,J}) results in $I_{ee} = 0$.
As it was shown in Refs. \cite{Nagaev08} and \cite{Nagaev09}, the main contribution to the current (\ref{dI}) comes from
collisions at points $\br$ located much farther from the gap than its size $a$.
Hence $\Omega_{in}(\br)$ may be considered as small and the fewer of the four momenta are in $\Omega_{in}(\br)$,
the larger the contribution to the current.
Also in Refs. \cite{Nagaev08} and \cite{Nagaev09} it was shown that
the largest contribution to the current
comes from the collisions of electrons incident on the gap
with electrons that are injected from the other half plane and have nearly opposite momentum.
Therefore we can assume that only
%for two momenta the domain of integration is limited to $\Omega_{in}(\br)$:
%and the incident one in $\Omega_{out}(\br)$.
%($\Omega_{out}(\br)$ contains all the momenta of electrons
%that will come from point $\br$ to the contact.
%It is obvious that at any point $\br$ domains $\Omega_{in}(\br)$ and $\Omega_{out}(\br)$ have the same area,
%but contain electrons with opposite momenta. If $\bk$ is in $\Omega_{in}(\br)$, $-\bk$ is in $\Omega_{out}(\br)$.)
%
the electrons with momentum $\bk$ are injected and lie in $\Omega_{in}(\br)$
while the electrons with the rest of momenta $\bp$, $\bp'$, and $\bk'$
are native to the considered half-plane.
%
%Therefore the integration over $\bk$ and $\bp$ in
%Eq. (\ref{dI}) may be limited to $\bk, -\bp \in \Omega_{in}(\br)$.

We sequentially integrate in Eq. (\ref{dI}) over the  time, coordinate and momenta as it was done in Ref. \cite{Nagaev09} for the case of low voltage $eV \ll T$. Here we consider the case of arbitrary voltages and obtain the correction to the current in a form of an integral over the energies
\begin{multline}
 I_{in} = \frac{e a^2 \alpha_{ee} m}{2(2\pi)^3}
\ln\frac{l_c}{a}\int{d\eps_{\bp}}\int{d\eps_{\bk}}\int{d\eps_{\bp'}} \int{d\eps_{\bk'}}
\\ \times
 F_0(\eps_{\bp}, \eps_{\bk}, \eps_{\bp'}, \eps_{\bk'})\,\delta(\eps_{\bp} + \eps_{\bk} - \eps_{\bp'} - \eps_{\bk'})
\\ \times
 \Theta(D)/\sqrt{D},
\label{dI_4eps}
\end{multline}
where $l_c$  is a cutoff length much larger than $2a$, which  may be due to a very weak electron-impurity scattering or a finite size of the electrodes,
\begin{multline}
 F_0(\eps_{\bp}, \eps_{\bk}, \eps_{\bp'}, \eps_{\bk'})
\\
 = [1 - f_{L}(\eps_{\bp})]\,
 [1 - f_{R}(\eps_{\bk})]\,
 f_{L}(\eps_{\bp'})\,f_{L}(\eps_{\bk'})
\\ {}
  -
  f_{L}(\eps_{\bp})\,f_{R}(\eps_{\bk})\,
[1 - f_{L}(\eps_{\bp'})]\,[1 - f_{L}(\eps_{\bk'})],
\label{F_L}
\end{multline}
and
\be
D = \left[(\eps_{\bp} - \eps_{\bk})^2 - (\eps_{\bp'} - \eps_{\bk'})^2\right]/4
\ee
is a value characterizing the deviation of the energies from the Fermi surface, which vanishes when all the energies lie exactly at the Fermi surface.
This expression allows us to analytically obtain the results in the limiting cases of high and low voltages and numerically calculate the correction for arbitrary relations $eV/T$. At high voltages $eV/T \gg 1$ the correction to the current has a form
\be
I_{in} = \frac{e a^2 \alpha_{ee} m}{(2\pi)^3}\,\ln\frac{l_c}{a} \times \left(1 - \frac{\pi}{4}\right) 
{\rm sign}(V) (eV)^2
\label{I-high}
\ee
and at low voltages $eV/T \ll 1$
\be
I_{in} = \frac{e a^2 \alpha_{ee} m}{(2\pi)^3}\ln\frac{l_c}{a} \times \frac{C_{10}}{2} (eV) T
\label{I-low-V}
\ee
where the constant $C_{10} = 3.72$.

Using the above semiclassical model, we can calculate the noise spectral density. It is expressed through the Fourier transform of the current correlation function as follows
\be
S = 2 \int\limits_{-\infty}^{\infty}{dt\, e^{i\omega t} \la\delta I(t)\,\delta I(0)\ra}.
\label{S}
\ee
We will calculate the spectral density at zero frequency $\omega = 0$.
Current fluctuation can be expressed in terms of fluctuation of the distribution function by the Eq. (\ref{current}), so the current correlator has a form
\begin{multline}
\la\delta I(t)\,\delta I(0)\ra =
e^2 \int\limits_{-a}^{a}{d\rho_1} \int\limits_{-a}^{a}{d\rho_2} \int{\frac{d^2p_1}{(2\pi)^2}} \int{\frac{d^2p_2}{(2\pi)^2}}
\\ \times
v_{1\perp}v_{2\perp}\,
\la\delta f(\bp_1, {\boldsymbol \rho_1}, t)\,\delta f(\bp_2, {\boldsymbol \rho_2}, 0)\ra.
\label{correlator_I}
\end{multline}
The fluctuation $\delta f(\bp, \br, t)$ obeys the Boltzmann-Lan\-ge\-vin equation \cite{Kogan}
\begin{align}
\left(
 \frac{\partial}{\partial t}
 + {\bf v}\frac{\partial}{\partial {\bf r}}
 + e{\bf E}\frac{\partial}{\partial {\bf p}}
\right)&
\delta f(\bp, \br, t)
+ \frac{\partial f}{\partial {\bf p}}\, e\delta {\bf E}
\nonumber\\ =
\delta I_{ee}(\bp, \br, t) +& \delta J^{ext}(\bp, \br, t),
\label{BL}
\end{align}
where $\delta J^{ext}(\bp, \br, t)$ is a Langevin source.
The correlator of Langevin sources was calculated  in \cite{Kogan69} on the assumption that each collision is correlated only with itself and equals
\begin{multline}
\la \delta J^{ext}({\br_1}, t_1, \bp_1)\,\delta J^{ext}({\br_2}, t_2, \bp_2)\ra
=
\frac{1}{2}\, V_{vol}\, \delta(\br_1 - \br_2)
\\
\times\delta(t_1 - t_2 )
\biggl[\delta_{\bp_1 \bp_2}\sum\limits_{\bp' \bk \bk'}(J_{\bp'\bk' \to \bp_1\bk} + J_{\bp_1\bk \to \bp'\bk'})
\\ {}+
\sum\limits_{\bp' \bk'}(J_{\bp'\bk' \to \bp_1\bp_2} + J_{\bp_1\bp_2 \to \bp'\bk'})
\\ {}
-2\sum\limits_{\bk \bk'}(J_{\bp_1\bk \to \bk'\bp_2} + J_{\bk'\bp_2 \to \bp_1\bk}) \biggr]
\label{JJ}
\end{multline}
We can neglect the field terms  in (\ref{BL}) for the same reason as it was done for the Boltzmann equation (\ref{Boltz}). As we consider finite temperatures,
we have to take into account equilibrium fluctuations of $f$ far from contact. To this end, we present $\delta f$ as a
sum of fluctuation $\delta f_0$ that has arrived from the depth of electrode by freely propagating without scattering
\cite{Kulik'}
and the integral of the right-hand side of Eq. (\ref{BL}) over time $\tau$ of travel to point $\br$ along the trajectory
of a free electron
\begin{align}
\delta f(\bp, \br, t) &= \delta f_0(\bp, \br, t)
+
\int\limits_{0}^{\infty}{d\tau} \left[\delta I_{ee}(\bp, \br - {\bf v}\tau, t - \tau) \right.
\nonumber\\ {}+&
\left. \delta J^{ext}(\bp, \br - {\bf v}\tau, t - \tau)\right].
\end{align}
Therefore the correlator of distribution functions in Eq. (\ref{correlator_I}) is given by
\begin{multline}
\la\delta f(\bp_1, {\boldsymbol \rho_1}, t)\,\delta f(\bp_2, {\boldsymbol \rho_2}, 0)\ra
\\
=
\la\delta f_0(\bp_1, {\boldsymbol \rho_1}, t)\,
\delta f_0(\bp_2, {\boldsymbol \rho_2}, 0)\ra
\\ {}
+
\int\limits_{0}^{\infty}{d\tau} \Bigl[\la\delta f_0(\bp_1, {\boldsymbol \rho_1}, t)\,
 \delta I_{ee}(\bp_2, \rho_2 - {\bf v_2}\tau, - \tau)\ra
\\ {}
+
\la\delta f_0(\bp_2, {\boldsymbol \rho_2}, 0)\,\delta I_{ee}(\bp_1, \rho_1 - {\bf v_1}\tau, t - \tau)\ra\Bigr]
\\ {}
+
\int\limits_{0}^{\infty}{d\tau_1}\int\limits_{0}^{\infty}{d\tau_2}\,
\la\delta J^{ext}(\bp_1, {\boldsymbol \rho_1} - {\bf v_1}\tau_1, t - \tau_1)
\\
{} \times
\delta J^{ext}(\bp_2, {\boldsymbol \rho_2} - {\bf v_2}\tau_2, - \tau_2)\ra.
\label{correlator_f}
\end{multline}
Note that $\delta f_0$ and $\delta J^{ext}$ are totally uncorrelated because of the causality principle.
The first term in (17) is the two-time correlation function of fluctuations that originate from the depth 
of electrodes and propagate to the point of observation without scattering. This correlation function is 
well known \cite{deJong96}
\begin{align}
\la\delta f_0&(\bp_1, {\boldsymbol \rho_1}, t)\delta f_0(\bp_2, {\boldsymbol \rho_2}, 0)\ra   =
(2\pi)^2\,\delta(\bp_1 - \bp_2)
\nonumber\\
\times &\delta({\boldsymbol \rho_1} - {\boldsymbol \rho_2} - {\bf v_1}t)\,f(\bp_1)[1 - f(\bp_1)].
\label{correlator_f0}
\end{align}
Substituting this correlator to (\ref{correlator_I}) and (\ref{S}) results in the Nyquist equation $S_0 = 4T G_0$, where $G_0$ is the Sharvin conductance.

To the first order in the scattering, the spectral density $S_{in} = S - S_0$  is defined by the last three summands in Eq. (\ref{correlator_f}).
In the second and the third summands of this equation, $\delta I_{ee}$ is obtained by variating the collision integral $I_{ee}$ (\ref{I_ee,J}) with respect to $\delta f$. To obtain the results to the first order in the interaction, we substitute  the distribution functions (\ref{f_zero}) in $\delta I_{ee}$ and evaluate the correlators $\la\delta f_0\,\delta I_{ee}\{\delta f\}\ra$ and $\la\delta I_{ee}\{\delta f\}\,\delta f_0\ra$ using the zero-approximation correlators Eq. (\ref{correlator_f0}). The fourth summand in Eq. (\ref{correlator_f}) is obtained directly from Eq. (\ref{JJ}). Then we substitute the resulting expressions into Eqs. (\ref{correlator_I}) and (\ref{S}).

\begin{figure}[t]
\includegraphics[width=0.95\linewidth]{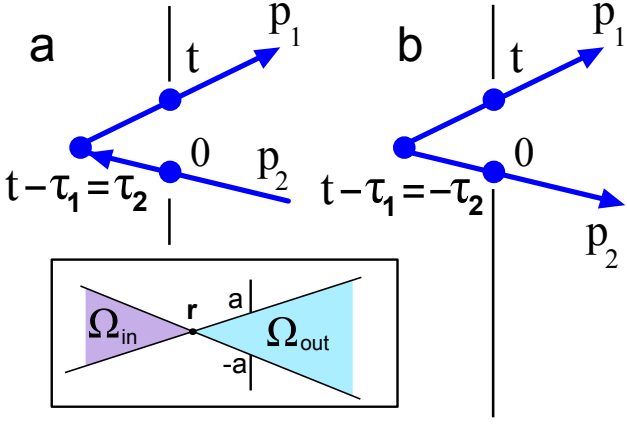}
\caption{Fig. 1. Illustration of the two terms in Eq. (\ref{S_flux_general}). An electron with momentum $\bp_1$  crosses the gap at time $t>0$. Another electron with momentum $\bp_2$  crosses the gap at time $0$. The collision takes place at $t-\tau_1<t$.
(a) The first term %in  Eq. (\ref{S_flux_general}) is proportional to $\delta(t - \tau_1 - \tau_2)$ and
corresponds to the collision of electrons between the crossing of the gap; $\tau_2$ is the time  between the first crossing  and the collision.
(b) The second term %in Eq. (\ref{S_flux_general}) is proportional to $\delta(t - \tau_1 + \tau_2)$ and
corresponds to the collision before both crossings; $\tau_2$ is the time between the collision and the first crossing.
The inset shows the domains $\Omega_{in}(\br)$ and $\Omega_{out}(\br)$.
}
\label{fig1}
\end{figure}

After some rearrangements the first-order spectral density  takes up a form
\begin{multline}
S_{in} = 2e^2 V_{vol} \int\limits_{0}^{\infty}{dt} \int\limits_{-a}^{a}{d\rho_1} \int\limits_{-a}^{a}{d\rho_2} \int{\frac{d^2p_1}{(2\pi)^2}} \int{\frac{d^2p_2}{(2\pi)^2}}
\\ {}\times
v_{1\perp}v_{2\perp} \int\limits_{0}^{\infty}\!{d\tau_1}
\int\limits_{0}^{\infty}\!d\tau_2\,
\delta({\boldsymbol \rho_1} - {\boldsymbol \rho_2} + {\bf v_2}\tau_1 - {\bf v_1}\tau_1 - {\bf v_2}t)
\\ {}\times\!\!
\left.\Bigl[
 \delta(t - \tau_1 - \tau_2)\, \Gamma_1 + \delta(t - \tau_1 + \tau_2)\, \Gamma_2
\Bigr]\right|_{\brho_1 -\bv_1\tau_1},
\label{S_flux_general}
\end{multline}
where $\Gamma_1$ and $\Gamma_2$ are combinations of distribution functions and scattering fluxes
\begin{multline}
\Gamma_1 =
-\sum\limits_{\bp' \bk'}
\Bigl\{
 f(\bp_2)\,J_{\bp'\bk' \to \bp_1 \bp_2}
\\
+ [1 - f(\bp_2)]\,J_{\bp_1\bp_2 \to \bp'\bk'}
\Bigr\}
+
2\sum\limits_{\bk \bk'}
\Bigl\{
 f(\bp_2)\,J_{\bp_1\bk \to \bp_2 \bk'}
\\
+ [1 - f(\bp_2)]\,J_{\bp_2\bk' \to \bp_1\bk}
\Bigr\},
\label{CL1}
\end{multline}

\begin{multline}
\Gamma_2 =
\delta_{\bp_1\bp_2}[1 - 2f(\bp_1)]\!
\sum\limits_{\bp' \bk \bk'}
(J_{\bp'\bk' \to \bp_2 \bk} - J_{\bp_2\bk \to \bp'\bk'} )
\\  {}+
[1 - f(\bp_1) - f(\bp_2)]
\sum\limits_{\bp' \bk'}(J_{\bp'\bk' \to \bp_1 \bp_2} - J_{\bp_1\bp_2 \to \bp'\bk'} )
\\ {}+
2\,[f(\bp_2) - f(\bp_1)]\sum\limits_{\bk \bk'}
( J_{\bp_1\bk \to \bp_2 \bk'} - J_{\bp_2\bk' \to \bp_1\bk}).
\label{CL2}
\end{multline}
The two terms in Eq. (\ref{S_flux_general}) have different physical meaning. The first of them corresponds to
the case where the collision takes place during the time interval between the two crossings of the gap by the participating electrons
(Fig. \ref{fig1}a). It originates from the collision integral in Eq. (\ref{BL}) and determines the Nyquist noise at $V=0$. This term vanishes at $T=0$ for any $V$ because it results from the equilibrium fluctuations
in the depth of electrodes. The second term corresponds to the case where both electrons cross the gap after the collision (Fig. \ref{fig1}b) and results from the corrections to the one-time correlation function of $\delta f$. In equilibrium this function is a thermodynamic quantity and does not depend on the strength of scattering. Therefore 
the scattering corrections to it and the second term in Eq. (\ref{S_flux_general}) vanish at $V=0$.

Consider now (\ref{S_flux_general}) and isolate the dominant terms in it. To give a contribution
to  $\delta I$'s, both the momenta $\bp_1$ and $\bp_2$ must lie either in the angular domain $\Omega_{in}$ or
in the centrally symmetric domain $\Omega_{out}$ (see Fig. \ref{fig1}, inset).
As well as for the correction to the current, the contribution
to the spectral density of noise is dominated by electron collisions far from the contact, so the angular domains are small and the contribution to the noise is maximum if a minimally possible number of electron momenta involved in scattering is restricted to $\Omega_{in}$ or $\Omega_{out}$. On the other hand, the colliding electrons must have almost opposite momenta to ensure maximum phase space available for the scattering.

In the first term with $\delta(t - \tau_1 - \tau_2)$, momenta $\bp_1$ and $\bp_2$ have opposite directions (see Fig. \ref{fig1}a) and lie in $\Omega_{out}$ and $\Omega_{in}$, respectively. To ensure maximum phase space for the
scattering, $\bp_1$ and $\bp_2$ must correspond either to the two initial or the two final states.
Hence only the first sum
%terms with  $J_{\bp_1\bp_2 \to \bp'\bk'}$ and $J_{\bp'\bk' \to \bp_1 \bp_2}$
should be retained in Eq. (\ref{CL1}).

In the second term with $\delta(t - \tau_1 + \tau_2)$, momenta $\bp_1$ and $\bp_2$ are both in $\Omega_{out}(\br)$ (see Fig. \ref{fig1}b) and cannot have opposite momenta. To make Eq. (\ref{CL2}) nonzero, the electron with $\bk$ must be
in $\Omega_{in}(\br)$. Hence the dominant contribution to Eq. (\ref{CL2}) arises from the first term
because $\delta_{\bp_1 \bp_2}$ in it lifts one of the three restrictions on the momentum integration.

We substitute the corresponding terms of Eqs. (\ref{CL1}) and (\ref{CL2}) into Eq. (\ref{S_flux_general}), sequentially integrate over times, momenta and coordinates and obtain
the spectral density in a form of an integral over energies

\begin{figure}[t]
\includegraphics[width=\linewidth]{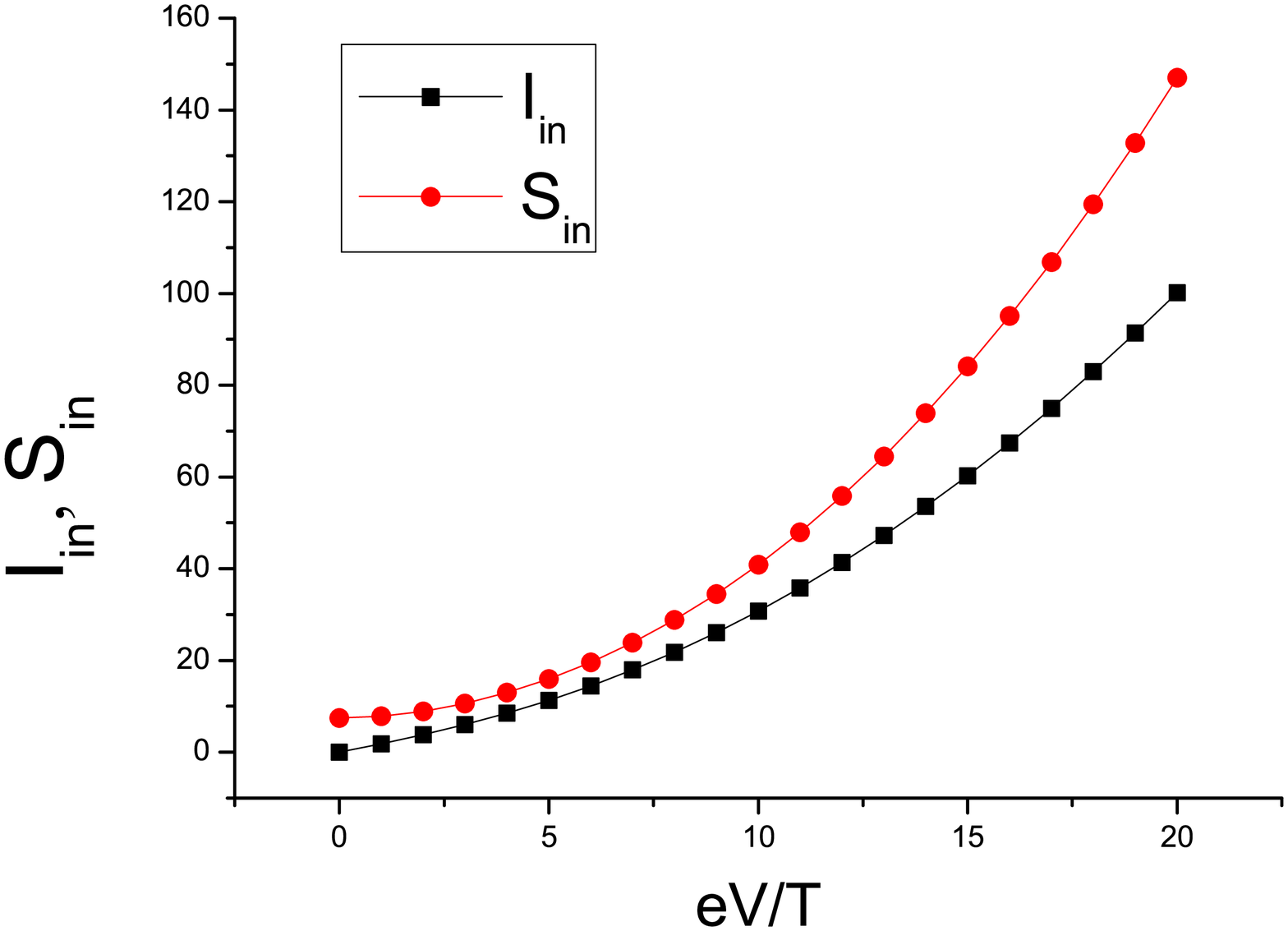}
\caption{Fig. 2. Dependencies of corrections to the current, measured in $ea^2\alpha_{ee}m/(2\pi)^3$, and spectral density, measured in $e^2a^2\alpha_{ee}m/(2\pi)^3$, on $eV/T$.}
\label{fig2}
\end{figure}
\begin{multline}
S_{in} = \frac{e^2 a^2 \alpha_{ee} m }{(2\pi)^3}\ln\frac{l_c}{a}
\int{d\eps_{\bp}} \int{d\eps_{\bk}} \int{d\eps_{\bp'}} \int{d\eps_{\bk'}}
\\ \times
\delta(\eps_{\bp} + \eps_{\bk} - \eps_{\bp'} - \eps_{\bk'})\,\Theta(D)\,D^{-1/2}
\\ \times
\bigl[
F_{1}(\eps_{\bp}, \eps_{\bk}, \eps_{\bp'}, \eps_{\bk'})
+ F_{2}(\eps_{\bp}, \eps_{\bk}, \eps_{\bp'}, \eps_{\bk'})
\bigr],
\label{S_4eps}
\end{multline}
where
\begin{multline}
F_{1} = 2 f_R(\eps_{\bk})[1 - f_R(\eps_{\bk})]
\Bigl\{ [1 - f_L(\eps_{\bp})]\,f_L(\eps_{\bp'})f_L(\eps_{\bk'})
\\ {}
+ f_L(\eps_{\bp})[1 - f_L(\eps_{\bp'})][1 - f_L(\eps_{\bk'})]
\Bigr\}
\end{multline}
and
\be
F_{2} = [1 - 2f_L(\eps_{\bp})]F_0
\ee
with $F_0$  defined by Eq. (\ref{F_L}).

At low voltages $eV \ll T$
\be
S_{in} = \frac{e^2 a^2 \alpha_{ee} m }{ (2\pi)^3}\ln\frac{l_c}{a} \times
\left[2C_{10}T^2 + C_2 (eV)^2 \right]
\ee
where the constants  $C_{10} = 3.72$ and $C_2 = 0.22$ were calculated numerically. In view of Eq.
(\ref{I-low-V}), this is in full agreement with the Nyquist theorem.

At high voltages $eV \gg T$ the spectral density has a form
\be
S_{in} = \frac{e^2 a^2 \alpha_{ee} m }{ (2\pi)^3}\ln\frac{l_c}{a} \times 2 \left(1 - \frac{\pi}{4}\right)(eV)^2
\ee
and is related with the inelastic contribution to the current (\ref{I-high}) by the Shottky formula $S_{in} =
2eI_{in}$. This is a consequence of the first approximation in the scattering and the fact that the inelastic
correction to the current is dominated by collisions far from the contact. The weak scattering suggests that
different electron collisions may be considered as independent random events whose contributions to the current
simply sum up. As the collisions take place far from the contact, the angular domain $\Omega_{in}$ is small and scattering of electrons within it may be disregarded. Therefore any scattering event changes the number of
electrons crossing the contact by unity and this results in the classical shot noise of the inelastic correction
to the current.

For arbitrary relations between voltage and temperature the $S_{in}(eV/T)$ and $I_{in}(eV/T)$ dependencies can
be obtained by numerically integrating Eqs. (\ref{dI_4eps}) and (\ref{S_4eps}). The resulting curves are shown
in Fig. \ref{fig2}.

To summarize, we have calculated the nonlinear correction  to the current and noise from electron-electron scattering for arbitrary relations between voltage $V$ and temperature $T$. Both quantities are dominated by electron collisions at distances from the contact much larger than its size and are positive for all $V$. This is markedly different from the case of impurity scattering, which results in a negative correction to the conductance and a correction to the noise that is negative at low voltages and positive at high voltages. At low voltages, the correction to the noise is determined by
thermal fluctuations that emerge from the depth of electrodes. At high voltages, it is determined
by random collisions of nonequilibrium electrons and is related with the nonlinear correction to the current by the classical Shottky formula.

%At low voltages $eV \ll T$ the correction to current is proportional to $V T$, the correction to conductance
%to $T$ and the spectral density to $T^2$.
%At high voltages $eV \ll T$ the correction to the current is proportional to $V^2\sign V$ and the spectral
%density, to $V^2$.
%At high voltages the noise is related with the inelastic correction to the current by the Shottky formula
%$S_{in} = 2e\,I_{in}$.

An experimental test of the Shottky relation for wide ballistic contact in high-mobility samples could additionally verify that  positive magnetoresistance and linearly increasing with temperature conductance observed in \cite{Renard08} are associated with electron-electron scattering at large distances from the contact.

This work was supported by Russian Foundation for Basic
Research, Grant No. 10-02-00814-a,
by the program of
Russian Academy of Sciences, and by Dynasty Foundation.


\begin{thebibliography}{99}
\bibitem{Kulik'}
This term was disregarded in Ref. \cite{Kulik84} because the authors addressed the zero-temperature
limit.
\end{thebibliography}
\end{document}